\documentclass[a4paper, 11pt]{article}
\usepackage[caption=false,font=normalsize,labelfont=sf,textfont=sf]{subfig}

\usepackage{natbib} 
\usepackage[left=2.05cm,
			right=2.05cm,
			top=2.05cm,
			bottom=2.05cm,]{geometry}
\usepackage[colorlinks]{hyperref}

\usepackage{amssymb}

\usepackage{amsmath}
\usepackage{authblk}
\usepackage{graphicx}

\hypersetup{
    colorlinks=true,
    linkcolor=red,
    citecolor=cyan,
}

\begin{document}

\title{Tree species classification at the pixel-level using deep learning and multispectral time series in an imbalanced context} 

\author[1,2]{Florian Mouret $^*$}
\author[2]{David Morin}
\author[2]{Milena Planells}
\author[1]{Cécile Vincent-Barbaroux}

\affil[1]{Université of Orléans, USC INRAE 1328 / P2E laboratory (Physiology, Ecology and Environment), BP 6759, rue de Chartres 45067 Cedex 2, Orléans, France.}
\affil[2]{CESBIO, Université de Toulouse, CNES/CNRS/INRAE/IRD/UT3-Paul Sabatier, 18, Avenue Edouard Belin, 31401 Toulouse, France.}

\maketitle
\begin{abstract}
This paper investigates tree species classification using Sentinel-2 multispectral satellite image time-series (SITS). Despite their importance for many applications and users, such mapping is often unavailable or outdated. The value of using SITS to classify tree species on a large scale has been demonstrated in numerous studies. However, many methods proposed in the literature still rely on a standard machine learning algorithm, usually the Random Forest (RF) algorithm. Our analysis shows that the use of deep learning (DL) models can lead to a significant improvement in classification results, especially in an imbalanced context where the RF algorithm tends to predict towards the majority class. In our case study in central France with 10 tree species, we obtained an overall accuracy (OA) of around 95\% and an F1-macro score of around 80\% using three different benchmark DL architectures (fully connected, convolutional and attention-based networks). In contrast, using the RF algorithm, the OA and F1 scores obtained are 92\% and 60\%, indicating that the minority classes are poorly classified. Our results also show that DL models are robust to imbalanced data, although small improvements can be obtained by specifically addressing this issue. Validation on independent in-situ data shows that all models struggle to predict in areas not well covered by training data, but even in this situation, the RF algorithm is largely outperformed by deep learning models for minority classes. The proposed framework can be easily implemented as a strong baseline, even with a limited amount of reference data. 

\end{abstract}





\section{Introduction}
Tree species identification is necessary for many applications related to forest monitoring. For example, species information is used in combination with allometric equations for carbon estimation, since carbon growth and storage is species dependent \citep{Wang_2006, Navar_2009, Henry_2013}. The proportion of tree species and species combinations is also used as an indicator of biodiversity and forest resilience \citep{Cavers_2014, Kacic_2022}. In addition, climate change is increasingly affecting the forests, with large-scale abiotic (fire, drought, windthrows) and biotic (insects and pathogens) disturbances impacting the species composition. Hence, there is an urgent need for updated tree species maps, e.g. for accurate analysis of tree health \citep{Mouret_2023_jstars} or to support reforestation decisions \citep{William_2021, HANEDA2023100882, Shovon_2024}. Unfortunately, such information is not always available, accurate or updated at fine scale and for large areas. Standard ground-based inventories are time-consuming and cannot be conducted on a yearly basis. For instance, in France, the last open dataset provided by IGN (Institut national de l'information géographique et forestière) is outdated since it was conducted between 2007 and 2018 (BD Forêt\textsuperscript{\textregistered } V2, \citep{IGN_2019}). In addition, National Forest Inventory (NFI) plots are not freely available and do not cover forests continuously.

The use of remote sensing data has been identified as a very efficient way to map tree species in a timely manner over large areas and with fine spatial resolution \citep{Fassnacht2016, HANEDA2023100882}. Sentinel-2 (S2) satellites, are increasingly used for such application \citep{Blickensdrfer2024, Liu_2024, VaghelaHimali2024}. Indeed, they provide multispectral images worldwide with fine spatial resolution (up to 10m) and low revisit time ($\sim$~5 days in Europe) \citep{DRUSCH201225}. This high spatio-temporal resolution can be used to monitor timely changes in vegetation cover, which is important to highlight the specific phenology of each tree species (or more generally other type of vegetation) \citep{KOWALSKI2020102172}. Some studies have shown the potential interest of using additional data, such as Sentinel-1 \citep{Blickensdrfer2024}, but it is not the focus of our analysis, mainly because Sentinel-1 data have much higher computational and storage costs from an operational perspective.

Standard frameworks for mapping tree species with in situ data often rely on the Random Forest (RF) algorithm \citep{Breiman2001}. For example, \cite{Fassnacht2016} observed in their review that the RF algorithm is one of the most used algorithms for such tasks, with other standard machine learning methods such as support vector machine (SVM). In more recent studies, the RF algorithm is still largely used, e.g., \cite{Persson_2018, Karasiak2021, Hemmerling2021, Blickensdrfer2024}. The popularity of RF with remote sensing data can be explained by several advantages: it is less prone to overfitting, requires less data for training, and is faster than deep learning (DL) methods. Moreover, the RF algorithm is more easily interpretable than other algorithms, which can be interesting depending on the task at hand (see for example \cite{Mouret_2023_jstars} in the case of forest health detection). 

Despite its advantages, the RF algorithm is known to be affected by imbalanced data \citep{More2017}, which is a common problem when dealing with tree species mapping since the different species are not naturally equally distributed. Moreover, it is also known that the RF algorithm can be outperformed by deep learning approaches to classify remote sensing data at large scale \citep{Bellet_2023}. In line with these points, a recent review has highlighted the shift towards deep learning models for tree species classification \citep{Zhong2024}. However, the methods reviewed focus on patch-level approaches, as in \cite{Bolyn2022}, where the patches are 400$\times$400 S2 pixels in size. In many cases, having ground data that correspond to the size of very large patches can be difficult (in the review made in \cite{Zhong2024}, the patch sizes range from 64$\times$64 to 500$\times$500 pixels). For instance, in our use case we had field plots of various size, many of them being of size 3$\times$3 S2 pixels. In addition, working with patches is more computational intensive than working with single pixel time series. The results obtained in \cite{Xi2021} have showed that pixel-level approaches can outperform patch-level approaches for tree species classification with S2 images. This is particularly important in mixed-species forests, such as those in our study area, where large patches can consist of different tree species.

Therefore, in this paper we propose to explore pixel-level deep learning strategies adapted to work with time series \citep{Wang2017} and compare their results with those obtained with the RF algorithm, which is still used as a standard classifier in many recent studies. Our use case focuses on a relatively small dataset, $\sim$ 4400 reference plots, which is nevertheless larger than the dataset used in many studies, such as \cite{Xi2021}. Moreover, the study area is the Centre-Val de Loire region of France and its surroundings, which corresponds to a large area (11 S2 tiles, 110000 km$^2$) with wide variety  of tree species and silvicultural practices. We also validated our results on an independent validation dataset for 4 key species (oak, pine, beech and chestnut), mainly to analyze the generalization of the mappings to unseen areas and minority species. Complementary to previous studies such as \citep{Xi2021}, some key steps are analyzed in more details in our study, in particular the imbalanced data problem and the hyperparameter choice of the different algorithms. Finally, the implemented DL models used with the open source iota2 Python library \citep{iota2_2016}, a processing chain for the operational production of land cover maps from remotely sensed image time series, have been made available: \url{https://framagit.org/fl.mouret/tree_species_classification_iota2}. In this respect, the proposed framework can easily be applied to other case studies.

\section{Study area and data}

This section present the study area, reference plots and satellite data used for our analysis.

\subsection{Study area}

\begin{figure}[h!]
    \centering
    \includegraphics[width=0.7\textwidth]{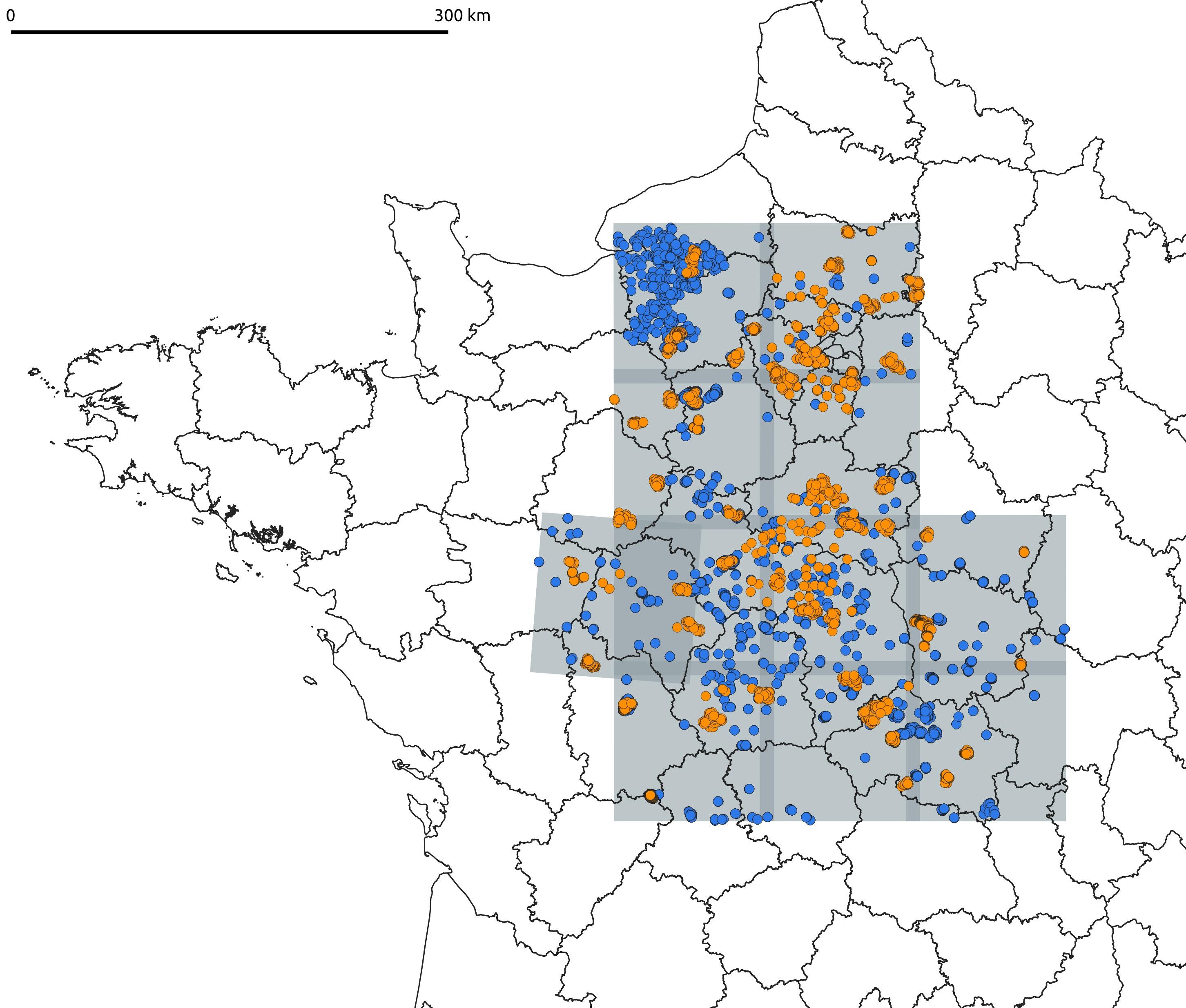}
    \caption{Our study area is delimited in grey, the boundaries between its 11 Sentinel-2 tiles is in lighter grey and administrative departments are outlined in black. The training plots are shown in blue, while the independent validation plots (4 species) are shown in orange.}
    \label{fig:study_area}
\end{figure}

Our study area is the Centre-Val de Loire region and its surroundings (11 S2 tiles, 110000 km$^2$), and is depicted in \autoref{fig:study_area}. The same area was analyzed in a previous study related to the detection of oak dieback, where the need for accurate mapping of tree species was identified \citep{Mouret_2023_jstars}. Our study area is a large region in northern France with a temperate climate and diverse forests. The region is a plateau with some hills, drained by the Loire River and its tributaries. Soil acidity and rainfall determine the type of forest. Thus, oak forests (65\% of the total \citep{IGN_2018}) with hornbeam, birch and chestnut dominate most of the region, which has drier and more acidic soils.

\subsection{Training  reference data}\label{sec:training_dataset}

\begin{figure}[h!]
    \centering
    \includegraphics[width=0.5\textwidth]{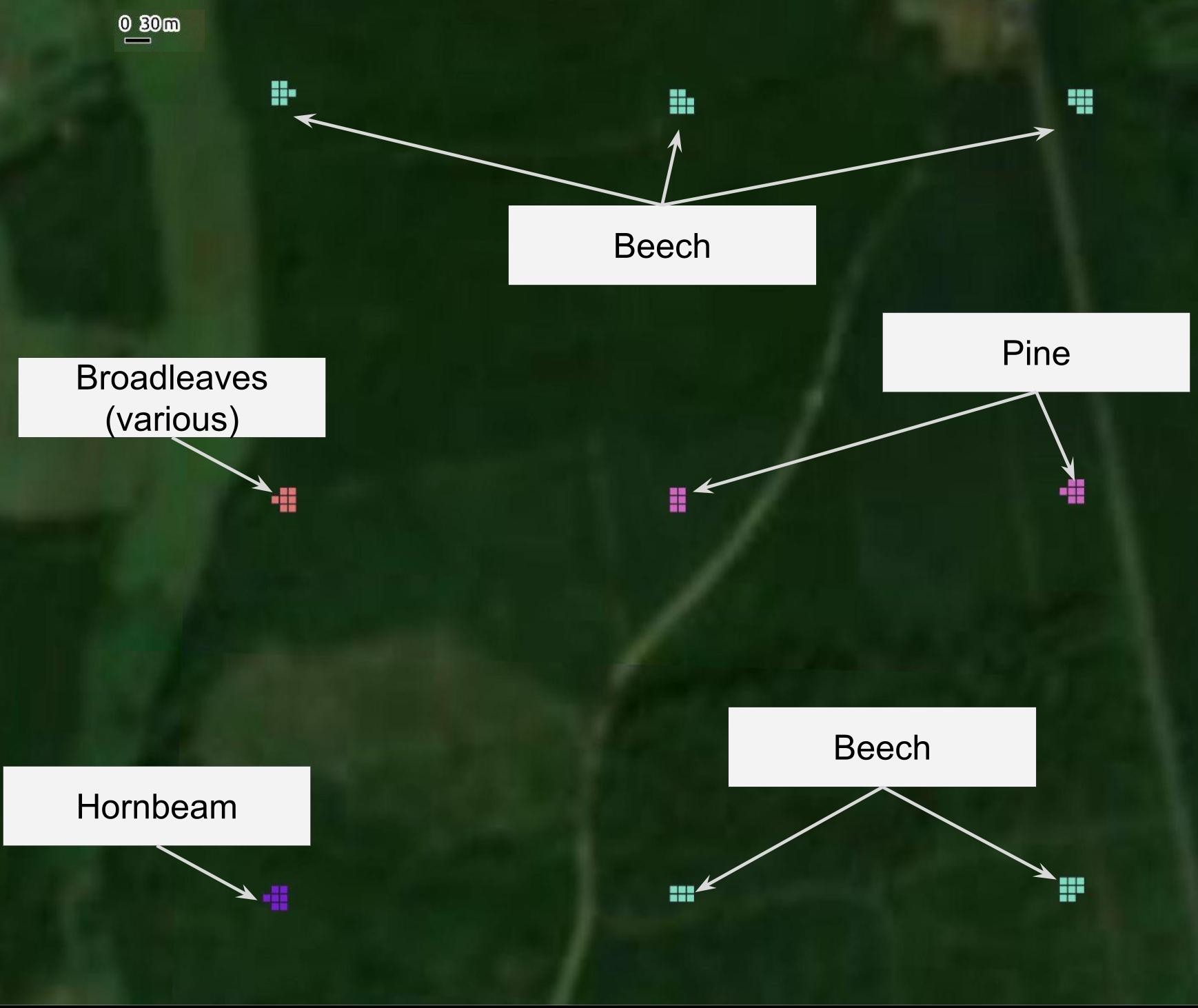}
    \caption{Illustration of different reference plots from our database. Each plot covers different S2 pixels (here the plots are very small, covering 6 to 8 pixels). Each dominant tree species is represented by a different color with the corresponding name.}
    \label{fig:ex_plot}
\end{figure}

Our reference data were provided by the ONF (Office National des Forêts). The reference plots are pure stands, i.e. more than 75\% of a single species 
(further work could be interesting to include plots with mixed species). The distribution of each species is provided in \autoref{tab:plots}. The total number of pixels is 67k (the ratio between the classes is about the same). It can be seen that oak plots dominate the data set, with more than 70\% of the plots belonging to this class (which is consistent with the fact that oak trees are predominant in our study area). The spatial distribution of the reference plots is neither systematic nor regular. \autoref{fig:study_area} shows a higher concentration of reference plots in the northwest of the study area, and some areas have no reference plots. Finally, \autoref{fig:ex_plot} provides an example of plots showing that 1) reference plots can be small and 2) different species can be found close to each other.

\begin{table}[h!]
\caption{Reference data used for our analysis.}
\centering
\begin{tabular}{llll}
Species & Definition & $\#$ Plots\\ \hline
Birch & Tree species of the genre \textit{Betula}      & 52   \\
Hornbeam & Tree species of the genre \textit{Carpinus}     & 48  \\
Chestnut & \textit{Castanea sativa Mill.}    & 61  \\
Oak & \textit{Quercus robur L.} and \textit{Quercus petraea (Matt.) Liebl.}   & 3219 \\
Douglas Fir & \textit{Pseudotsuga menziesii (Mirb.) Franco.}     & 131  \\
Fraxinus & Tree species of the genre \textit{Fraxinus}    & 39 \\
Beech &  \textit{Fagus sylvatica L.}   & 254  \\
Poplars &  Tree species of the genre \textit{Populus}   & 78  \\
Pines &  Tree species of the genre \textit{Pinus}   & 486 \\
Robinia &   \textit{Robinia pseudoacacia L.}  & 20 
\end{tabular}
\label{tab:plots}
\end{table}

\subsection{Additional independent validation data}\label{sec:validation_dataset}

We were able to validate our mappings using independent plots of 4 key species from forest health campaign assessments (these plots were used, for example, in \cite{Mouret_2023_jstars, Mouret_2024_IGARSS} to map forest dieback). In total, we had at our disposal 1497 oak, 197 chestnut, 91 pine and 136 beech pure plots from these campaigns. Having these independent plots is interesting since our training data is 1) highly imbalanced and 2), does not cover all the massifs of the region.

\subsection{Satellite data}

As explained in the introduction of this paper, our mapping is based on S2 data only, as we focus on a large area and want to propose a classification framework that can be easily used for similar tasks. We used the same processing chain as in \cite{Mouret_2023_jstars}, consequently a brief description is given below, more details are available in that reference. The S2 satellites (S2-A and S2-B) are operated by the European Space Agency for the European Union's Copernicus Earth observation program \citep{DRUSCH201225}. S2 data are multispectral images, we used 10 spectral bands for our analysis to cover the visible (bands 2, 3, 4), red-edge (bands 5, 6, 7), near infrared (NIR) (bands 8 and 8a) and short-wave infrared (SWIR) (bands 11 and 12) parts of the spectrum. Each image was resampled to pixels of size $10\times10$m. The MAJA processing chain \citep{Hagolle2015} was used to produce ortho-rectified level-2A images, with cloud and shadow masks. Finally, S2 images acquired between 2019 and 2020 were used, providing long-term canopy information (the effect of changing the temporal length of the input data is discussed in \autoref{sec:discussion}).

\section{Methods}

\begin{figure}[h!]
    \centering
    \includegraphics[width=0.5\textwidth]{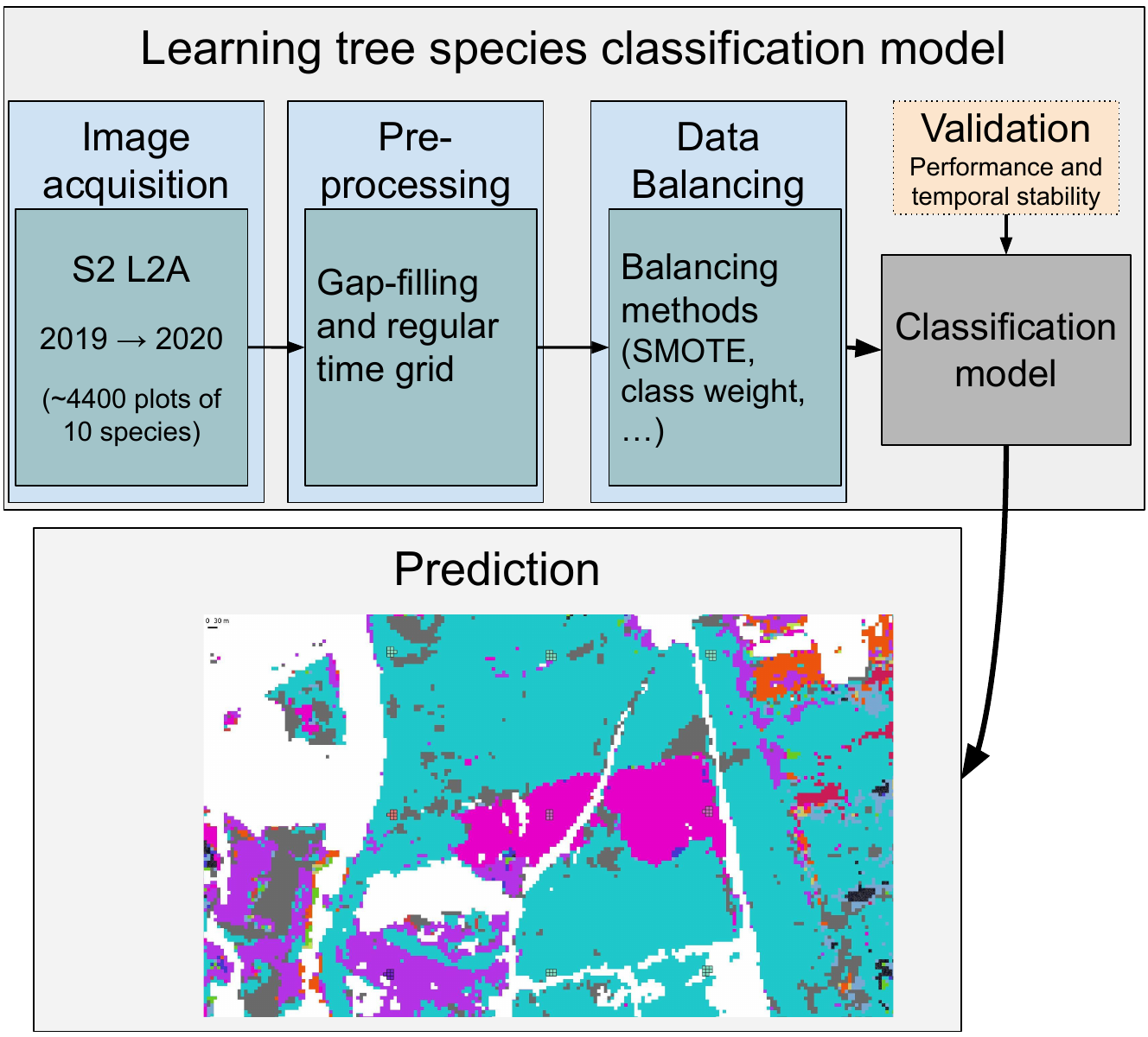}
    \caption{Methodological steps used to map tree species with S2 time series.}
    \label{fig:workflow}
\end{figure}

This section provides the methodological steps used to map tree species. A simplified workflow is depicted in \autoref{fig:workflow}. The $iota^2$ Python processing chain \citep{iota2_2016} was used to extract training samples and produce maps of our study area. As explained in the previous paragraph, level L2A S2 images are acquired and cloud-filtered. Then, they are interpolated on a regular time grid to produce 740 features for each pixel (10 bands over 74 dates). Finally, a classification algorithm is trained on the features of the reference data and used to produce a large-scale tree species map. More details on each step are given below.

\subsection{Preprocessing}

A linear interpolation (i.e., gap-filling) was used to have consistent time series for each S2 tile (1 interpolated data every 10 days) \citep{Inglada_2016, VUOLO2017202, Mouret_2023_jstars}. At the end of these preprocessing steps, each pixel is characterized by 740 features corresponding to 10 bands acquired over 2 years (74 dates). Other strategies for dealing with missing data and irregularly sampled time series were also tested (without improving our results) and are discussed in \autoref{sec:discussion}.

In order to be used efficiently by the classification algorithm (this is particularly true for the convolutional- and attention-based DL models), it is common to have standardized data with mean equal to 0 and standard deviation equal to 1. In our case, especially for the convolutional operation, it is important to preserve the temporal correlation of the S2 bands \citep{Pelletier2019}. Therefore, each S2 band was standardized by taking all the acquisitions over time (and not by standardizing each acquisition independently).

\subsection{Classification algorithms}
\subsubsection{RF algorithm}

Among the many classification algorithms available in the literature, the RF algorithm is widely used due to its scalability, robustness, and ability to model complex phenomena. The RF algorithm is an ensemble method based on decision trees: it constructs many decision trees and averages their predictions by majority voting. The RF algorithm uses bootstrap aggregation (known as bagging) to improve the stability and reduce the variances of the classification: each tree is trained on a subset of the original dataset, and at each split, a random subset of the features is selected. As explained in the introduction of this paper, we used the RF algorithm as a benchmark method because it is one of the most popular classification approaches in remote sensing. A major advantage of the RF algorithm is its robustness to the choice of its hyperparameters. For our experiments, we used default values, i.e. 100 fully grown decision trees (our tests have confirmed that changing these hyperparameters does not significantly affect the classification results).

\subsubsection{Deep learning models}

\begin{figure}[h!]
    \centering
    \subfloat[MLP]{\includegraphics[width=0.8\textwidth]{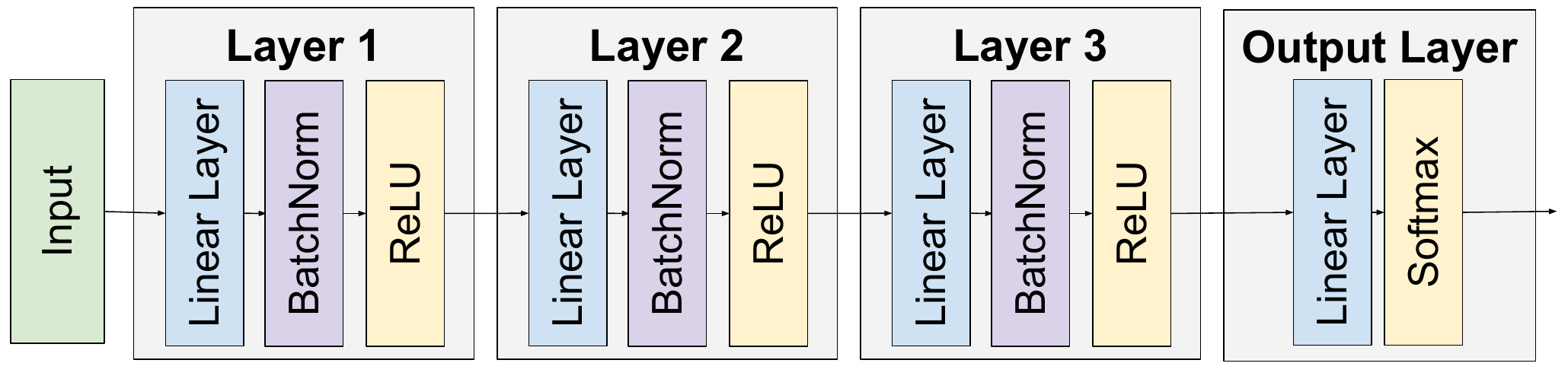}}

    \subfloat[TempCNN]{\includegraphics[width=0.8\textwidth]{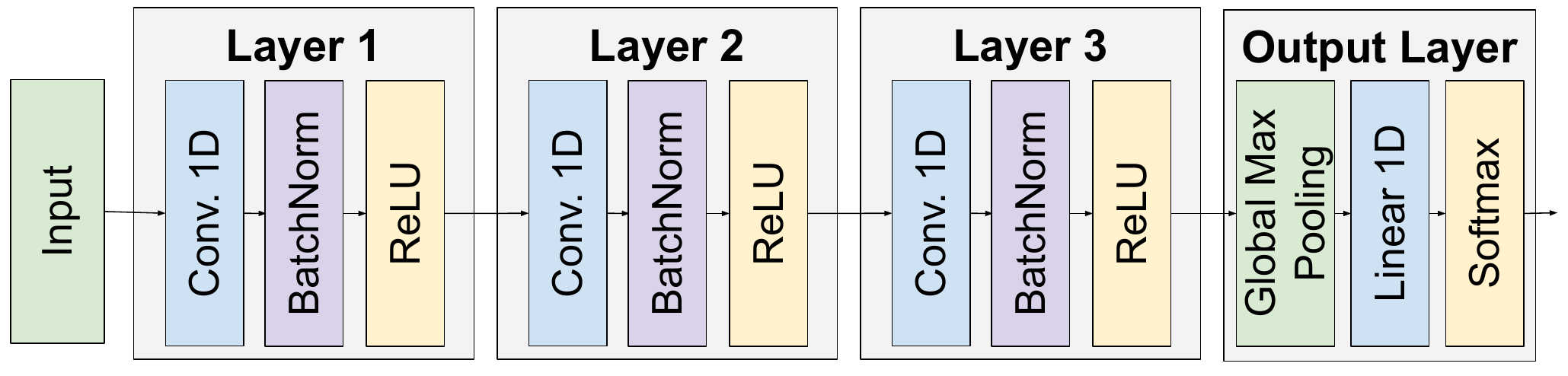}}

    \subfloat[LTAE]{\includegraphics[width=0.8\textwidth]{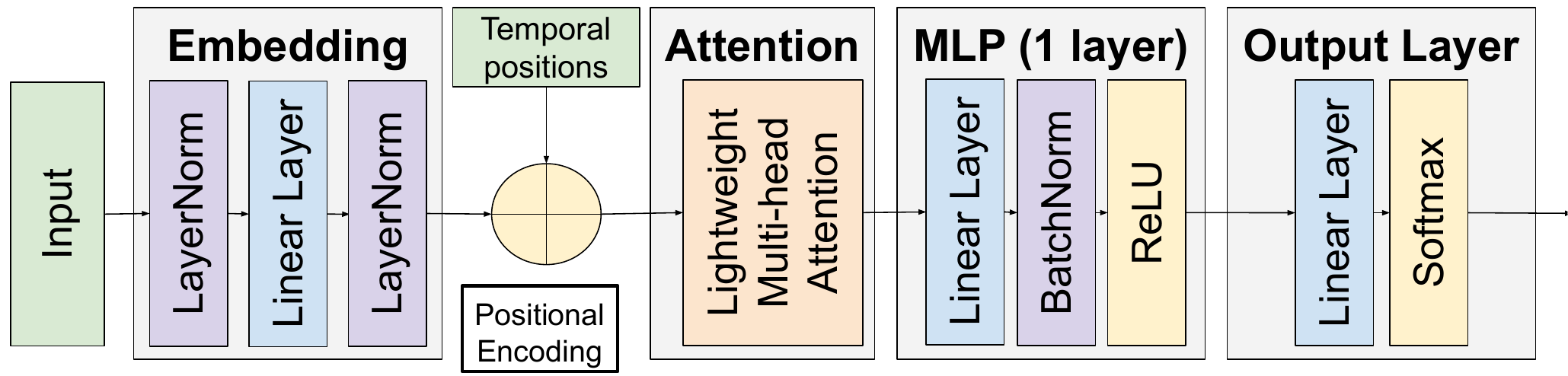}}
    \caption{The different deep learning architecture tested in our analysis. The last layer is a standard linear layer with a number of neurons equal to the number of classes to be predicted and a softmax activation.}
    \label{fig:DL_structure}
\end{figure}

In our analysis, we focused on 3 different DL architectures based on different mechanisms. We used two classical architectures, which are strong baselines to classify time series \citep{Wang2017}, namely multilayer perceptron (MLP or fully connected neural network) and temporal convolutional network (TempCNN)\citep{Pelletier2019}. These two models are easy to implement, and our results will show that they can work without extensive tuning, which is interesting for operational purposes. These standard architectures are depicted in \autoref{fig:DL_structure}: it consists of a succession of layers (we have fixed the number of layers to 3, as it provided the best results), each layer being composed of a transformation layer (i.e., a linear or convolutional layer), a batch normalization, and a nonlinear activation function (the rectified linear unit (ReLU) was used). Since we are working with time series, note that the convolution layers are 1-dimensional convolution layers. Unlike in \cite{Wang2017}, the MLP network also uses batch normalization instead of dropout, as it was found to be more efficient and stable. We also implemented a more recent method, the Lightweight Temporal Attention Encoder (LTAE) \citep{Garnot_2020}, which is based on the attention mechanism and is a state-of-the-art method for land cover classification \citep{Bellet_2023}. 

In summary, from a simplified point of view these 3 structures are very similar: they all aim at extracting relevant features, which are used by the output linear layer to classify the time series. The main difference is related to the feature extraction mechanism (dense layers, convolutions with global max pooling or attention mechanism).

\subsubsection{Hyperparameters used for the deep learning models}

Training a deep learning network is known to be more complex than classical machine learning algorithms such as RF. In the following, we propose a simple guideline for the choice of each parameter. The values used in our experiments are reported in \autoref{tab:params}. Our results show that these defaults parameters have produced good results for each model. We used the ADAM optimizer, other optimizers were tested and did not change our results significantly. The number of epoch was fixed to 100. An adaptive learning rate strategy was used to reduce the learning rate by a factor of 2 if the validation loss did not improve after 20 epochs. Moreover, the training was stopped after 40 epochs without improvements. The batch size was set to 2048, apart for the MLP with SMOTE where in that case it was set to 8192 (smaller values lead to very poor results, see our discussion on that point). Finally, the Cross-entropy loss was used, other losses were tested (e.g., margin loss) without improving our results.

\begin{table}[h!]
\caption{Hyperparameter values used in the different DL models.}
\centering
\begin{tabular}{llll}
 & MLP & TempCNN & LTAE \\
\hline
Learning rate & $1^{-4}$ & $1^{-3}$ & $1^{-3}$ \\
Neurons / conv. filters & 1024, 512, 256 & 128, 128, 128 & 512 \\
Filter size & - & 3,3,2 & - \\
Heads & - & - & 6 \\
Embedding size & - & - & 370 \\
Learnable parameters  & 1.4M & 113K & 271K
\end{tabular}
\label{tab:params}
\end{table}

The MLP and TempCNN can be easily tuned by setting a sufficient number of neurons/filters and an appropriate learning rate. Since the MLP tends to overfit more easily, we have found that reducing its learning rate to $1^{-4}$ instead of $1^{-3}$ was more stable. In particular, the use of numerous neurons (1024 or 2048) or filters (128) was found to be efficient even when the number of training samples is relatively small (see discussion in \autoref{sec:discussion}). For the MLP, we divided the number of neurons by 2 at each layer to reduce the size of the model and improve performances. The LTAE is slightly more complex to tune (number of heads, embedding size, final MLP). Setting the number of heads to 6 was the most effective in our case; reducing this number can lead to underfitting, and increasing it can lead to convergence problems (this may, of course, depend on the data set at hand). The dimension of query/key vector was set to 8 as in the original paper. Similarly, the embedding size was set to 370 (number of features divided by 2). As for the other models, we have found that setting the number of neurons in the MLP to a relatively high value (512 instead of 128 in the original paper) was found more efficient.

\subsection{Methods for dealing with imbalanced data}

Most of the methods used to deal with class imbalance can be grouped into algorithm-level methods (i.e., the training algorithm directly takes the class imbalance into account) and data-level methods (i.e., a modification of the training dataset is done)\citep{Johnson2019}. Our analysis mainly focuses on one method from each family. The first approach is an algorithmic-level method and consists in using class weights to penalize errors related to the underrepresented classes during training. The second strategy, which belongs to the data-level methods, consists in generating synthetic samples from the minority classes. We have used the classical algorithm Synthetic Minority Oversampling Technique (SMOTE), which has been widely used for a wide range of applications \citep{HaiboHe2009}. Finally, we also trained our DL models on our raw dataset without considering the class imbalance problem.

Note that other methods have been tested and a discussion of their performance is provided in \autoref{sec:discussion}. In particular, we tested a variant of SMOTE, namely ADASYN \citep{He_2008}, without any improvement in our results (see the additional results provided in \autoref{sec:results_additional}).

\subsection{Validation experiments and metrics}

Our experimental results have been validated by stratified cross-validation (CV, 10 folds). The train/test separation is done at the plot level to avoid auto-correlation problems. The main metric used to validate our results is the F1 macro score, i.e., the average F1 score of each class, a metric robust to imbalanced data. More precisely, the F1 score is the harmonic mean of precision and recall, where precision is the percentage of samples correctly labeled in class $j$ and recall is the percentage of samples in class $j$ that were correctly labeled. In addition, we also show the overall accuracy (OA), which is the percentage of samples correctly classified (so it is affected by class imbalance), and the balanced accuracy (BA), which is the recall averaged over each class (so it is not affected by class imbalance). These two additional metrics can be useful since their interpretation is intuitive. 

\section{Results}

\subsection{Classifier comparison and main results}

The CV results (F1, OA and BA) obtained with the different tested configurations are displayed in \autoref{tab:CV_main}. Overall, it is clear that the DL approaches largely outperform the RF algorithm, especially in terms of F1 and BA, which means that the RF algorithm struggles to correctly classify the underrepresented classes. Moreover, all DL models can achieve very close classification metrics, regardless of the strategy used to deal with imbalanced data. The MLP provides the best F1 score (0.81) without needing any strategy to deal with imbalanced data. Finally, note that the best BA (0.83) is obtained using LTAE with SMOTE (0.82 is obtained using MLP with SMOTE), which means that the recall of minority classes is higher on average.

\begin{table}[h!]
\caption{F1, OA and BA obtained for different classification configurations after 10-fold cross-validation. The 95\% confidence interval is in parentheses. Best values are in bold.}
\centering
\begin{tabular}{lllll}
 Model & Imb. strat. & F1 & OA & BA \\
 \hline
RF & - & 0.59 (0.06) & 0.93 (0.01) & 0.52 (0.05) \\
RF & Class weight & 0.62 (0.06) & 0.93 (0.01) & 0.54 (0.05) \\
RF & SMOTE & 0.62 (0.06) & 0.93 (0.01) & 0.54 (0.05) \\
MLP & - & \textbf{0.81 (0.03)} & \textbf{0.96 (0.01)} & 0.80 (0.04) \\
MLP & Class weight & \textbf{0.81 (0.03)} & \textbf{0.96 (0.01)} & 0.80 (0.04) \\
MLP & SMOTE & 0.80 (0.03) & 0.95 (0.01) & 0.82 (0.03) \\
TempCNN & - & 0.79 (0.03) & 0.95 (0.01) & 0.79 (0.03) \\
TempCNN & Class weight & 0.80 (0.02) & 0.95 (0.01) & 0.80 (0.02) \\
TempCNN & SMOTE & 0.80 (0.02) & 0.95 (0.01) & 0.79 (0.02) \\
LTAE & - & 0.80 (0.02)  & 0.95 (0.01)  & 0.77 (0.03) \\
LTAE & Class weight & 0.80 (0.04)  & 0.95 (0.01) & 0.81 (0.04) \\
LTAE & SMOTE & 0.80 (0.04) & 0.95 (0.01) & \textbf{0.83 (0.04)}
\end{tabular}
\label{tab:CV_main}
\end{table}

\begin{figure}[h!]
    \centering
    \subfloat[RF (SMOTE)]{\includegraphics[width=0.455\textwidth]{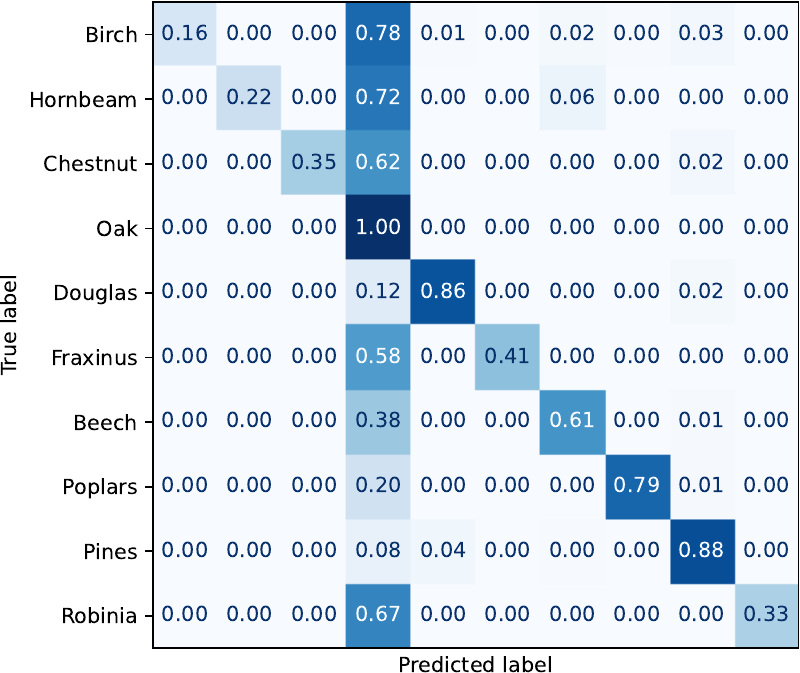}}~\subfloat[MLP (SMOTE)]{\includegraphics[width=0.45\textwidth]{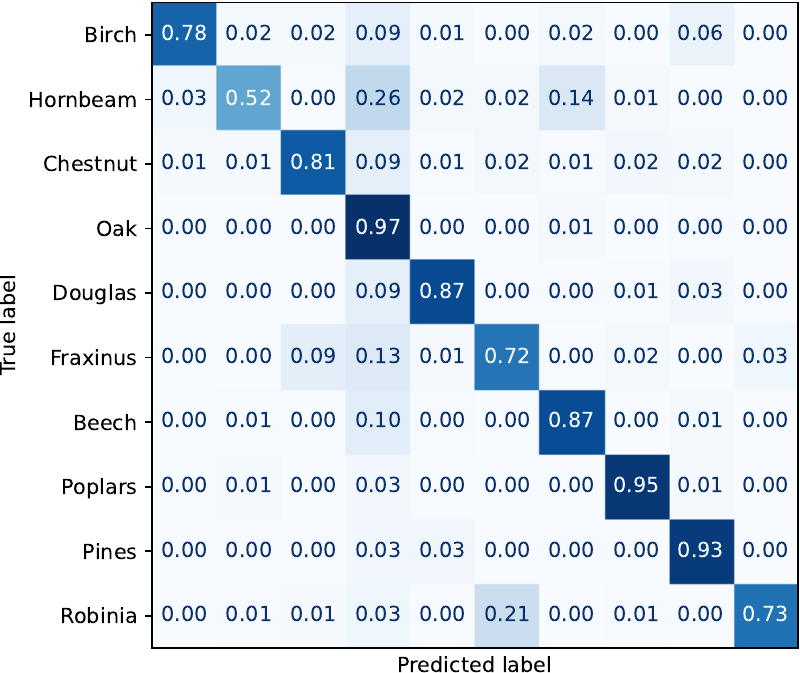}}

    \subfloat[TempCNN (SMOTE)]{\includegraphics[width=0.45\textwidth]{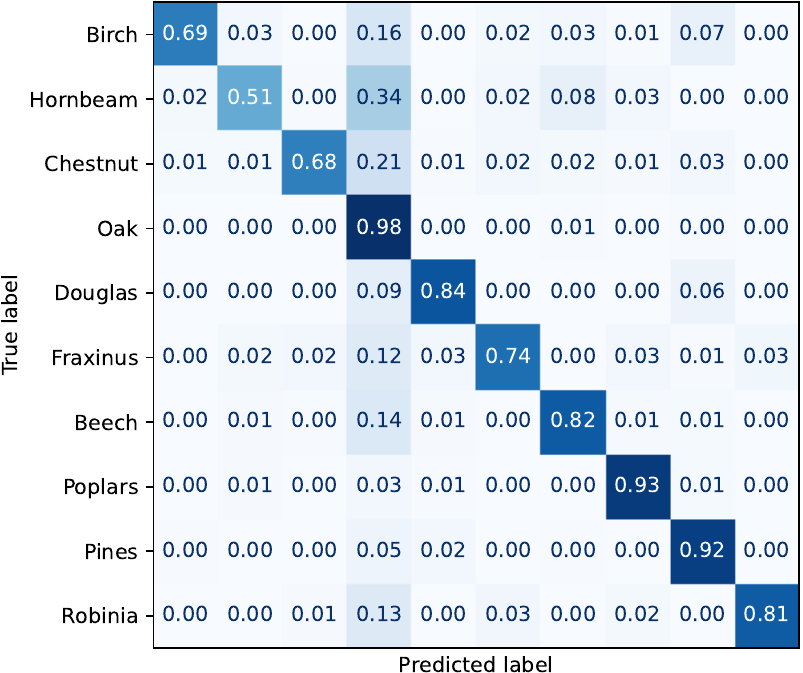}}~\subfloat[LTAE (SMOTE)]{\includegraphics[width=0.45\textwidth]{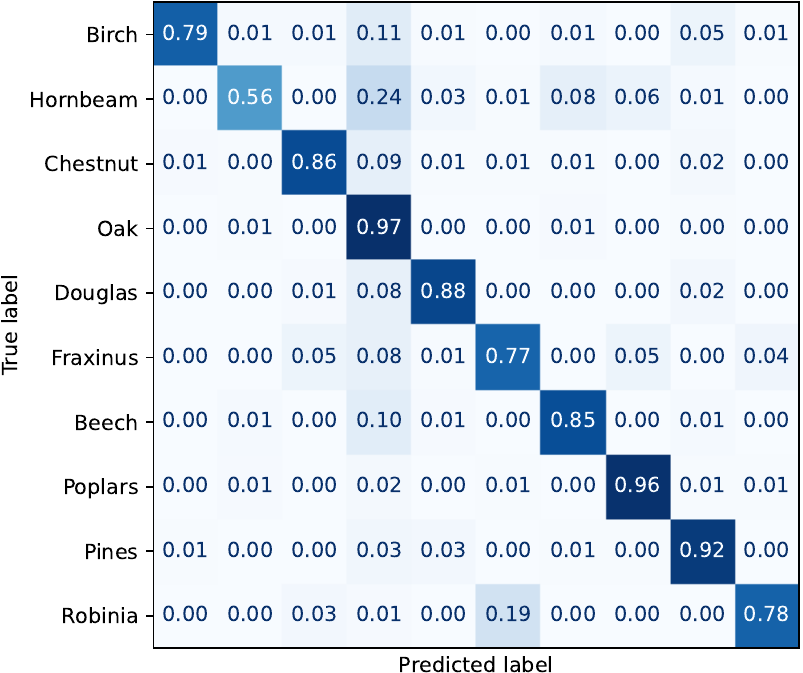}}

    \caption{Normalized confusion matrices averaged after 10 folds CV using SMOTE oversampling.}
    \label{fig:CM_all}
\end{figure}

In addition to these table, we have provided normalized confusion matrices averaged after the 10 runs for each classifier using SMOTE oversampling in \autoref{fig:CM_all} (similar results are obtained with other strategies). Looking at these confusion matrices, it is clear that the RF algorithm tends to classify the minority species (except conifers, poplars and beech) as oak, which explains the poor results in terms of F1 score and BA obtained in \autoref{tab:CV_main}. A significant improvement is observed with the DL methods, even if some minority species (such as birch, hornbeam, beech, and Fraxinus) can be partially classified as oak (between 10 and 30\% confusion, depending on the species and model). In addition, DL models also tend to confuse hornbeam with beech and robinia with fraxinus, tree species that are more similar to each other than to oak. Finally, the models with the best BA (typically LTAE and MLP with SMOTE) tend to have higher recall in minority classes (e.g. the best recall for 5 minority species is obtained with LTAE). However, the recall for oak tends to be slightly lower in this case (0.97 instead of 0.98 or 0.99), which affects the precision for other species as the dataset is highly imbalanced.


\subsection{Validation on independent data}

The RF, MLP, TempCNN and LTAE models were validated on our independent datasets described in \autoref{sec:validation_dataset}. Each model was trained using on the dataset presented in \autoref{sec:training_dataset}. For each species (oak, pine, chestnut and beech) we computed the percentage of plots correctly classified by the different models (i.e., recall). For this experiment, a plot is considered correctly classified if more than 50\% of its pixels are correctly classified (results obtained with other classification rules are discussed below). These results are reported in \autoref{table:valid}. 

Overall, most confusion was observed for chestnut and beech species, which tend to be predicted as oak. LTAE with SMOTE performs best and can retrieve on average 71\% of the plots, with a significantly higher recall for beech compared to other models (38\% for LTAE instead of 29\% for MLP, the second best model for this species). Very close results are obtained using MLP with SMOTE, which was found to be better for chestnut retrieval. Overall, it appears that using the SMOTE algorithm leads to better validation results compared to using the class weight, indicating better generalization to areas without training data. The RF algorithm is again largely outperformed, with a BA of 57\% and poor retrieval of chestnut and beech (which is consistent with the results obtained with the standard CV in the previous section).


We also perform the same analysis by considering that a plot is correctly classified when more than 20\% of its pixels have been correctly detected (this is an easier task since only a few pixels need to be correctly detected). In this case, with SMOTE oversampling, the average recall (BA) is 0.66, 0.85, 0.77, and 0.82 for RF, MLP, TempCNN, and LTAE, respectively. This significant improvement shows that at least a part of the pixels are generally correctly detected, which is valuable for real-world applications. Finally, pixel-level results were also computed and are very close to the results given in \autoref{table:valid} (BA equal to 0.58, 0.72, 0.67, and 0.69 for RF, MLP, TempCNN, and LTAE).

\begin{table}[h!]
\caption{Percentage of validation plots correctly retrieved (i.e., 50\% of the pixels are correctly classified).}
\centering
\begin{tabular}{lllllll}
Model & Imb. strat.   & Oak           & Pine          & Chestnut      & Beech         & Mean (BA)            \\
\hline
RF & SMOTE       & \textbf{0.99} & 0.92          & 0.33          & 0.06           & 0.58           \\
MLP & Class weight      & 0.98 & 0.91 & \textbf{0.62} & 0.24 & 0.69  \\
MLP & SMOTE      & 0.97          & 0.95 & 0.60 & 0.29 & 0.70 \\
TempCNN & Class weight      & 0.98 & 0.91 & 0.42 & 0.18 & 0.62 \\
TempCNN & SMOTE & 0.97 & 0.93 & 0.56 & 0.26 & 0.68 \\
LTAE & Class weight & 0.98 & 0.87 & 0.43 & \textbf{0.38} & 0.67 \\
LTAE & SMOTE     & 0.96 & \textbf{0.96} & 0.52 & \textbf{0.38} & \textbf{0.71}       
\end{tabular}
\label{table:valid}
\end{table}

\subsection{Additional results}\label{sec:results_additional}

To complement the previous results, \autoref{table:additional_results} provides additional experiments conducted to evaluate some variations in the classification workflow.

\begin{table}[h!]
\caption{F1 and BA obtained for different configurations after 10-fold CV. TempCNN is abbreviated CNN. ``Same'' means that the configuration is the same as in \autoref{tab:CV_main}. ``BS'' stands for batch size and ``BN'' for batch normalization. ``Imb. Strat.'' is the strategy used to deal with imbalanced data, ``Under.'' means that the oak plots were undersampled by randomly selecting 400 training plots. For SMOTE and ADASYN, the number of neighbors is given (SMOTE k=100 means SMOTE with 100 neighbors). The 95\% confidence interval is in parentheses.}
\centering
\begin{tabular}{llllll}
Model    & Config. & Imb. Strat.       & Data               & F1 & BA         \\
\hline
MLP      & Same   & Class weight       & 1 year  & 0.78 (0.04) & 0.78 (0.04) \\
MLP      & 128, 64, 32   & Class weight        & Same        & 0.80 (0.05) & 0.80 (0.04) \\
MLP      & 128, 64, 32   & -       & Same        & 0.79 (0.05) & 0.77 (0.04) \\
CNN & 64, 64, 64    & -       & Same        & 0.76 (0.03) & 0.70 (0.04) \\
LTAE & 4 heads       & -       & Same        & 0.79 (0.03) & 0.80 (0.04) \\
LTAE     & 8 heads       & -      & Same        & 0.75 (0.04) & 0.78 (0.03) \\
MLP     & BS 256      & -      & Same        & 0.80 (0.03) & 0.77 (0.03) \\
MLP     & BS 256      & Class weight      & Same        & 0.80 (0.03) & 0.79 (0.03) \\
MLP     & BS 8k      & Class weight      & Same        & 0.80 (0.03) & 0.80 (0.04) \\
MLP     & BS 8k no BN      & Class weight      & Same        & 0.73 (0.05) & 0.80 (0.04) \\
MLP     & No BN      & -      & Same        & 0.79 (0.02) & 0.77 (0.02) \\
RF       & Same   & Class weight       & Under. & 0.70 (0.03) & 0.68 (0.04) \\
MLP      & Same   & Class weight        & Under. & 0.69 (0.03) & 0.80 (0.04) \\
CNN & Same   & -       & Under. & 0.72 (0.02) & 0.77 (0.03) \\
MLP      & BS 2048   & SMOTE, k=5  & Same        & 0.73 (0.03) & 0.82 (0.03) \\
MLP      & BS 4096   & SMOTE, k=5  & Same        & 0.77 (0.03) & 0.82 (0.03) \\
MLP      & Same   & SMOTE, k=100  & Same        & 0.79 (0.03) & 0.81 (0.03) \\
MLP      & Same   & ADASYN, k=5   & Same        & 0.78 (0.03) & 0.79 (0.03) \\
MLP      & Same   & ADASYN, k=100 & Same        & 0.79 (0.02) & 0.81 (0.03) \\
CNN & Same   & SMOTE, k=100  & Same        & 0.80 (0.02) & 0.80 (0.02) \\
CNN & Same   & ADASYN, k=5   & Same        & 0.76 (0.03) & 0.72 (0.03) \\
CNN & Same   & ADASYN, k=100 & Same        & 0.80 (0.01) & 0.79 (0.02)
\end{tabular}
\label{table:additional_results}
\end{table}

\begin{itemize}
    \item Time range used for the analysis: using 1 year of S2 data instead of 2 slightly decreases the classification results (F1=0.78 instead of 0.81), we observed this with the other models.
    \item Size of the DL models: regarding the size of the DL models, a good F1 score can be obtained with an MLP with 3 layers of 128, 64 and 32 neurons, with or without class weights. Note that without class weight, the BA is lower (0.77 instead of 0.80). Overall, we have found that smaller network tend to be more impacted by imbalanced data. In practice, we have observed that choosing a larger number (typically 1024 or 2048 neurons in the first layer) is not a problem and can lead to a slight improvement in classification, which is interesting because it can be adapted to larger datasets. We found that at least 128 filters were needed for TempCNN, with a drop in accuracy observed when using 3 layers of 64 filters. On the other hand, we observed that adding layers was not beneficial and led to a slight deterioration of our classification results. For LTAE, using too many heads can lead to lower accuracy due to convergence problems. However, using 4 heads instead of 6 results in a slightly lower BA (0.80 instead of 0.83).
    \item Regularization: a small batch size (here, 256) can reduce the accuracy of the results. In our experiment, when using a batch size equal to 256 with an MLP, even if the F1 score is close to the optimal value, we observed a decrease in the BA (0.77 instead of 0.80). In this case, using class weights mitigates this problem (0.79 instead of 0.80). Using a very large batch size (e.g., 8192) was not found to affect the overall results when using batch normalization (without it, F1 score drops to 0.73 when using a batch of size 8192). Finally, without batch normalization, the F1 score drops to 0.79 instead of 0.81 (and the BA to 0.77 instead of 0.80). The same conclusions were found with or without class weights, and also with the TempCNN architecture.
    \item Other strategies to deal with imbalanced data: our results show that undersampling the oak plots (here, by randomly selecting 400 plots) improves the RF predictions (which are still far from the optimal scores) but worsens the F1 score obtained with DL methods, implying that overall it is better to use all available samples with an appropriate strategy to deal with imbalanced data. Regarding the oversampling strategies, SMOTE (S) and ADASYN (A) can lead to very similar results. Regarding the choice of their hyperparameters, we have found that the number of neighbors used in the SMOTE algorithm does not have much impact on the classification results, while it is important to choose a large number (e.g., 100 or more) in the ADASYN algorithm. Finally, when using MLP with SMOTE, we observed a significant reduction in the F1 score when reducing the batch size (0.73 instead of 0.80 when using a batch size of 2048).
\end{itemize}

\section{Discussion and perspectives}\label{sec:discussion}

The fact that the RF algorithm can be affected by imbalanced data has already been identified in the literature \citep{chen04using, More2017, Blickensdrfer2024}. In \cite{Mellor2015}, in a land cover classification context, the authors highlighted that RF tends to predict towards the majority class, which is consistent with our observations. In our case, we have found that using RF with SMOTE or class weight failed to handle this issue. As observed in \autoref{sec:results_additional}, undersampling the oak plots improves the results of the RF algorithm, but the F1 score we obtained is still much lower than those obtained with DL methods. For example, this strategy was used in \cite{Blickensdrfer2024} to map tree species in Germany, where the authors found that mapping less common species was still challenging, confirming the results we obtained with the RF algorithm. We also tested a variant of the RF algorithm, namely the Balanced RF \citep{chen04using} implemented in the Python toolbox imbalanced-learn \citep{Lemaitre_2017}. However, this variant lead to very poor results. As a key takeaway, we recommend that practitioners consider these potential issues for future work, especially since our results highlight the potential benefit of using DL methods, which we found to be able to generalize better, especially for less frequent classes. Other classification methods such as Sparse Gaussian Processes (\citep{Bellet_2023}), Support Vector Machine (SVM) algorithm \citep{Cortes1995} or Tree Boosting Algorithms (XGboost) \citep{Chen_2016} have also been tested without providing competitive results. In particular, we also tested other DL architectures (e.g., ResNet, 2D convolutional neural network, recurrent neural network, etc.) without improving our classification results. This means that most of the DL architectures can achieve similar accuracy for our task.

Our results highlight the potential benefits of using DL models, and some simple guidelines can be followed to achieve good accuracy. Specifically, choosing a model sufficiently large (e.g., in our case at least 200k parameters), applying batch normalization and class weights is a good start. Furthermore, using an MLP with these specifications can provide a strong baseline model that requires minimal hyperparameter tuning. On the basis of these recommendations, a review of the literature revealed that there was a tendency for some studies to use DL models with too few parameters. For example, \cite{Verhulst2024} implemented an MLP with only 1 hidden layer with 100 neurons for tree species classification, which led to poor results. Similarly, \cite{Zagajewski2021} used a single layer with 18 hidden units. In both cases, the small number of neurons could explain the poor performances obtained with DL methods.

Regarding the method used to address data imbalance, slight improvements and better generalization can be achieved by oversampling with the SMOTE algorithm, but in this case we found that choosing an appropriate batch size was important. Indeed, we observed that if the SMOTE algorithm is used with too small a batch size, poor results can be obtained. This obervation is especially true for the MLP, which is probably due to the fact that MLP can easily overfit, hence they are sensitive to small batches that can be composed of synthetic noisy samples. Finally, the LTAE architecture was found to give the best performance when using the SMOTE algorithm, although it required more hyperparameter tuning. Additional tests using different losses (e.g., margin or focal loss, label smoothing) or the ADASYN algorithm did not improve our results. Overall, it appears that DL methods can also learn patterns related to the minority classes if the model is sufficiently large, hence they can be robust to the imbalanced data problem. This is consistent with previous studies (conducted for other types of data) that showed that good tuning of the models can be sufficient to achieve optimal accuracy in an imbalanced data context \citep{shwartz-ziv2023simplifying}.


Regarding the input data used for classification, we observed that the direct use of the raw S2 band is sufficient. Similarly to \cite{Mouret_2023_jstars}, we observed that using 2 years of data instead of 1 can be useful to better characterize the forests and thus improve the classification results. Using additional sensors and information, such as synthetic aperture radar (SAR) \citep{PRIYANKA2023100924, Blickensdrfer2024} or hyperspectral data \citep{DMITRIEV2023100964}, is an interesting perspective that could improve our results.

We have also tested other strategies to deal with missing data (related to clouds). Our results confirm that standard gap-filling (i.e., linear interpolation) is a strong baseline adapted to work at large scale (11 S2 tiles) and in an operational context \citep{Inglada_2015, Bellet_2023, Mouret_2023_jstars}. However, we observed that poor interpolation could impact the generalization of the results, especially for minority classes, since few training examples are available. To that extent, working on this point is an interesting perspective, recent studies have shown that improvement is possible by working directly with raw (irregular and unaligned) SITS \citep{Bellet_2024}. Our initial tests were not very conclusive: we tested the method developed in \cite{Bellet_2024}, but this lead to inaccurate classification results. This approach, tested for land cover classification, proposes an end-to-end learning framework that makes use of an attention-based interpolation to embed the raw SITS into a regular temporal grid. More tests should be done to investigate why such an approach fails in our case (imbalance data, small dataset, classes with very similar behavior, etc.). We also conducted experiments using the LTAE with raw S2 data and a cloud mask, as tested in \cite{Bellet_2024}. While theoretical results obtained with standard CV appears close to those obtained with our proposed framework, our tests at large scale obtained by producing a map of the study area have shown important changes at the frontier of two S2 tiles. This highlights the fact that such framework tend to largely overfit the specific acquisitions of a given S2 tile, which confirm the results observed in \cite{Bellet_2024}.

Other interesting perspectives could be explored, three of which are developed below. We observed that combining the different maps produced with the different models, e.g., by taking the mode of the predictions, could filter some noisy areas in some cases. However, our first tests show that the gain in classification accuracy is not always obvious, hence further work is needed to develop an optimal ensemble learning approach. Nevertheless, we observed that the three different DL architectures produced varying predictions in certain areas, which could serve as an indicator of mapping uncertainty. Interestingly, using different initializations within the same DL architecture yielded significantly more consistent results, providing little to no basis for uncertainty measurement. Finally, for operational mapping, it may be interesting to add more reference data to produce maps that are as accurate as possible. Using accessible information on the tree species from BD Forêt\textsuperscript{\textregistered } V2, \citep{IGN_2019} could be an obvious way for increasing the size of the training dataset. However, since this database is not up-to-date, adding such samples might be non-trivial. Working on mixed-species plots could be another way to add more training samples to the dataset, the main problem being to automatically separate these plots into pure class pixels.

\section{Conclusion}

This paper investigates tree species classification in temperate forests based on multispectral remote sensing data in an imbalanced context and over a large area (110000 km$^2$). Our study area is located around the Centre-Val de Loire region of France, which is dominated by oak forests (75\% of our training plots), and our training dataset is relatively small (less than 5000 plots). The proposed framework uses 2 years of Sentinel-2 (S2) data as input features (a preprocessing is done to remove clouds and interpolate the different input time series on the same temporal grid).

Our results highlight that deep learning (DL) models can be used in that context with good accuracy and largely outperform classical machine learning model such as the Random Forest algorithm, which is highly biased toward the majority class in our case. We tested 3 different architecture based on different mechanisms, i.e., a fully connected network (MLP), a convolutional-based network (1D convolutional network, TempCNN) and an attention-based network (LTAE). These implementations and configuration files have been made freely available and can be used as baseline for other use cases (and potentially other applications). 

Our results show that the 3 deep learning architectures tested provide similar results and largely outperform the RF algorithm. The MLP architecture was found to be a strong baseline (F1 score equal to 0.81) that can be used without extensive hyperparameter tuning. While more complex to implement, the LTAE also provided good generalization capabilities and appears to be more efficient at retrieving minority classes when used with the SMOTE algorithm, an oversampling technique. Overall, we found that using the SMOTE algorithm with an appropriate batch size could slightly improve the generalization of the classification models.

\section*{Data availability} 

The configuration files used to train the models and map the study area with the iota2 processing chain are available here: \url{https://framagit.org/fl.mouret/tree_species_classification_iota2}. The open-source iota2 project can be used to produce land cover or vegetation maps from remotely sensed time series, more information and implementation can be found here: \url{https://framagit.org/iota2-project/iota2}.

\section*{Acknowledgments}
The authors would like to thank the M. Fauvel, H. Touchais and all the TOSCA PARCELLE team for his help and expertise on the use of iota2 software. We thank ONF for sharing reference data and CNES for the use of its HPC center.

\section*{Author contributions} Conceptualization, F.M, D.M., M.P. and C.V-B.; methodology, F.M, D.M., M.P. and C.V-B.; software, F.M.; validation, F.M, D.M., M.P. and C.V-B.; formal analysis, F.M, D.M., M.P. and C.V-B.; investigation, F.M.; resources, F.M.; data curation, F.M. and D.M; writing---original draft preparation, F.M.; writing---review and editing, F.M, D.M., M.P. and C.V-B.; visualization, F.M.; supervision, M.P. and C.V-B.; project administration, M.P. and C.V-B.; funding acquisition, M.P. and C.V-B. All authors have read and agreed to the published version of the manuscript.

\section*{Fundings} This work was supported by the SYCOMORE program, with the financial support of the Région Centre-Val de Loire (France), in collaboration with the SuFoSaT project of the GRAINE ADEME program.
 

\bibliographystyle{elsarticle-harv} 
\bibliography{references}
\end{document}